\documentclass{article}
\usepackage[utf8]{inputenc}
\usepackage[lmargin=0.65in,rmargin=0.65in,tmargin=0.9in,bmargin=0.75in]{geometry}
\usepackage{hyperref,lmodern,textcomp,graphicx,float}
\usepackage[table]{xcolor} \usepackage{caption,fancyhdr}
\usepackage{color,amsmath,amssymb,mathpazo,mathtools,bbm,babel}
\usepackage{vwcol}
\usepackage{lipsum}
\usepackage{tgpagella}
\usepackage{indentfirst}
\usepackage{bchart}
\usepackage{lettrine,multicol}

\usepackage{tcolorbox}%
\definecolor{mycolor}{rgb}{0.122, 0.435, 0.698}% Rule colour
\makeatletter
\newcommand{\mybox}[1]{%
  \setbox0=\hbox{#1}%
  \setlength{\@tempdima}{\dimexpr\wd0+13pt}%
  \begin{tcolorbox}[boxrule=0pt,arc=0pt,
      left=6pt,right=6pt,top=6pt,bottom=6pt,boxsep=1pt,width=\textwidth]
    #1
  \end{tcolorbox}
}

\newcommand{\HBA}[1]{{\color{black} #1}}

\fancypagestyle{firstpage}
{\fancyhead[L]{\textsc {\color {gray} {\large }}}    
\fancyhead[R]{\large \textsc{\color {gray} {Letter}}}}
    
\graphicspath{{Figures/}}

\begin{document}

\thispagestyle{firstpage}
{\noindent \LARGE\textit{Elastic Lattices Inspired by Ulam-Warburton Cellular Automaton}} \\ [0.5em] 

\noindent {\large \textit{Hasan B. Al Ba'ba'a}} \\

\noindent \begin{tabular}{c >{\arraybackslash}m{6in}}
    \includegraphics[]{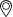} &
    \noindent {\small Department of Mechanical Engineering, Union College, Schenectady, NY 12308, USA} \\[0.25em]
    
    \includegraphics[]{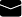}& \noindent {\href{mailto:albabaah@union.edu}{\small albabaah@union.edu}}\\
    
\end{tabular}

%%%%%%%%%%%%%%%%%%%%%%%%%%%%%%%%%%%%%%%%%%%%%%%%%%%%%%%%%%%%%%%%%%%%%%%%%%%%%%%%%%%%%%%%%%%%%%%%%%%%%%%%%%%%%%%%%%%%%%%%%%%%%%%%%%%%%%%%%%

\mybox{Periodic lattices have been widely explored for decades, owing to their peculiar vibrational behavior.~On the other hand, certain types of aperiodic lattices have enabled new phenomena that may not be otherwise attainable in periodic ones. In this paper, a new class of aperiodic lattices inspired by cellular automaton is introduced. Cellular automata were originally developed as a machine replication algorithm and it has been intensively explored in computer science.~These algorithms yield structures that are not necessarily periodic, yet follow well-defined rules that lead to interesting patterns.~The concept is utilized here to build elastic lattices following such rules, and Ulam-Warburton Cellular Automaton (UWCA) is demonstrated as an example.~Starting from a square monatomic lattice, an UWCA lattice is constructed and its vibrational behavior is analyzed, showing unique dynamical properties, including symmetric eigenfrequency spectra, repeated natural frequencies of large multiplicity, and the emergence of strongly localized corner modes.~It is envisioned that computer-algorithm-inspired lattices may unlock new wave phenomena that could outperform existing lattice designs.}

\vspace{0.25cm}
\noindent \textbf{Keywords
}

\noindent Cellular Automata, Ulam-Warburton, Natural Frequencies, Mode Multiplicity, Corner Modes, Aperiodicity.

\noindent\rule{7.2in}{0.5pt}
\begin{multicols}{2}
%\section{Introduction}

Lattices, a particularly interesting class of elastic periodic systems, are defined as structures built from simple interconnected elastic elements such as rods, beams, and plates.~Due to their relatively lightweight and unique mechanical properties~\cite{Pan2020,Askari2024}, lattices have enabled a variety of phenomena in wave propagation domain, such as wave directionality \cite{phani2006wave}, energy harvesting \cite{Gonella2009}, cloaking \cite{nassar2020polar}, and zero-energy modes~\cite{lubensky2015phonons,xia2021microtwist}.~On the lattice design front, optimization techniques have been applied to effectively engineer lattices with frequency ranges associated with blocked wave propagation (i.e., bandgaps \cite{sigmund2003systematic,bilal2011ultrawide}) and topological protection \cite{du2020optimal}.~Furthermore, fractal-inspired designs of lattices has gained popularity and proven to exhibit interesting phenomena.~As a case in point, lattices with a Koch fractal design have demonstrated control over bandgaps by varying the number of fractal iterations, such that the frequency at which the first bandgap opens decreases as the iteration number increases~\cite{zhao2020elastic}.~Another study highlights the direct influence of fractal design on altering the elastic constant tensor, which in turn leads to different elastic response and generated bandgaps, where the latter is greatly affected by the fractal-cut symmetry of the unit cell~\cite{kunin2016static}.~Fibonacci-based structures have also been investigated, yielding interesting patterns in their eigenfrequencies/bandgaps, which have been correlated with the Fibonacci sequence~\cite{hou2004acoustic,bacigalupo2022design}. Finally, bioinspired designs (comprising nacre, spider webs, cochlea, moth wings, and fractal structural shapes) have recently garnered research interest, and numerous studies in this domain have been encapsulated in a review by Dal Poggetto~\cite{dal2023bioinspired}.

\begin{figure*}[t]
     \centering
\includegraphics[]{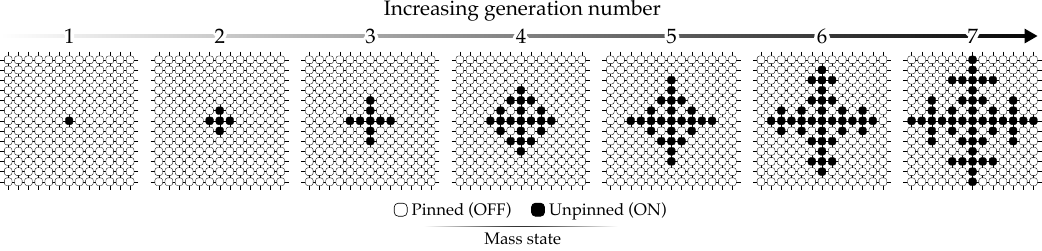}
\caption{\textbf{Elastic lattices inspired by Ulam-Warbutron Cellular Automaton (UWCA):} The first seven generations of the considered UWCA lattice are depicted.~The initial (zeroth) generation starts as an infinite square monatomic elastic lattice with pinned  masses $m$ (or in an ``OFF" state), interconnected via springs of stiffness $k$.~The first generation is enabled by unpinning the central mass, which is interpreted here as the ``ON" state.~Subsequent generations are determined by unpinning masses that share a spring connection with exactly one unpinned (or an ``ON") mass, as dictated by the UWCA algorithm.}
     \label{fig:Schematic}
\end{figure*}

While periodic lattices have been a focal point of research for a few decades, the notion of aperiodicity and exploring new ways for controlling waves is becoming increasingly appealing.~For instance, quasi-periodicity realizes topologically protected edge states with various edge modes' localization \cite{rosa2021exploring,gupta2020dynamics}.~Additionally, the concept of topological pumping, where a two-dimensional structure is adiabatically modulated to gradually pump a signal from one corner to an opposite one, has been experimentally and theoretically demonstrated \cite{riva2020edge,rosa2019edge}.~Quasi-periodicity has also enabled the concept of near zero modes, known as floppy modes, which constitute a class of topological modes in Maxwell lattices \cite{zhou2019topological}.~Analogously, the concept of rainbow trapping, achieved by an array of resonators with a change in the system parameters that are not necessarily periodic (e.g., Ref.~\cite{de2021selective}), finds application in areas such as vibration isolation \cite{zhang2022rainbow} and energy harvesting \cite{wang2021metamaterial}.

Building upon the promise of lattices that are unique in design and not necessarily periodic, and motivated by the need for novel approaches to control elastic/acoustic waves, I take inspiration from Cellular Automata (CA) to introduce a new class of lattice designs.~By definition, CA are discrete computational models, invented by John von Neumann in the 1950s~\cite{li2018theory}, originally to study self-replicating systems. Beyond computer science, CA have since been widely adopted across various fields such as biology, physics, and engineering.~For instance, a biological object can be considered a quantifiable entity by numbers, and these numbers can relay how the different objects can self-replicate and influence each other.~Given their role as a universal computational methodology for machine self-replication \cite{sarkar2000brief,kari2005theory}, CA are leveraged to computationally model and understand such biological entities and living organisms, such as protein, RNA, and DNA \cite{chaudhuri2018new}.~In physics, CA have been instrumental in studying many systems, including (i) the behavior of sand piles subject to a blow of wind \cite{cattaneo2012sand}, (ii) computational modeling of traffic flow problems \cite{maerivoet2005cellular}, and (iii) statistical mechanics \cite{wolfram1983statistical}.~More pertinent studies to this work are (i) the use of hybrid CA, a computational method originally developed to understand the structural bones adaptation \cite{tovar2006topology}, to optimally design piezoelectric patches anchored to vibrational skin for maximizing its power output \cite{lee2013topology}, (ii) the characterization of fracture behavior in elastic bodies under dynamical loading via continuous-discontinuous CA \cite{yan2018continuous}, and (iii) implementing CA as a computational modeling approach for wave propagation~\cite{Kluska2013Lamb_CA,NISHAWALA2016_CA_Experiment}, nonlinear Ultrasonic waves~\cite{autrusson2009numerical}, and elastodynamic responses of seismic events and other loading scenarios~\cite{LEAMY2008seismic_elastodynamics_CA,Hopman2010Elastodynamic_CA}.

For the sake of illustration, I demonstrate here a lattice design based on Ulam-Warburton Cellular Automaton (UWCA), named after the contributions of Ulam and Warburton~\cite{warburton2019ulam}.~Rather popular in mathematical circles~\cite{ulam1962some,applegate2010toothpick,warburton2019ulam}, the UWCA algorithm can be conceptualized using an infinite sheet of packed square cells.~These square cells exist in a binary state, typically denoted as “ON” or “OFF” (may also be interpreted as ``1" or ``0" or ``dead" or ``alive" cells).~Initially, a pristine set of squares is set at an “OFF" state, except for a single central square which is designated ``ON". The defining rule of UWCA mandates that the neighboring ``OFF" squares turn ``ON" \textit{only} if they share a single edge with an ``ON" square.~If an ``OFF" square cell shares two or more edges with ``ON" cells, that square remains inactive (i.e., ``OFF").~Every stage the pattern grows is called a generation~\cite{ulam1962some} (assigned the variable $n_g$ henceforth), and this growth is a topic of mathematical analysis to predict the growth after a known number of generations \cite{applegate2010toothpick}.~An additional rule for the considered UWCA is that if an ``OFF" cell becomes ``ON" at any given generation, it stays ``ON" throughout all future generations.~Note that UWCA rules can be applied to other regular shapes, like a hexagon or triangle \cite{ulam1962some}.

%\section{Results}

%\subsection{UWCA elastic lattice}

\HBA{To demonstrate UWCA in elastic lattices, I start by a square monatomic elastic lattice, constituting identical masses $m$ interconnected by taut strings under tension and exhibit out-of-plane motion. The mathematical model of the lattice is adopted from Hussein \textit{et al.}~\cite{hussein2014dynamics}, where the interconnecting elements are approximated by linear springs with an equivalent spring constant $k$ (the reader is referred to Al~Ba'ba'a \textit{et al.}~\cite{AlBabaa2020Elastically-supportedInsulators} for more information regarding this approximation).}~Implementing the rules of UWCA requires defining ``OFF" and ``ON" states within the context of a vibratory lattice.~To this end, the ``OFF" state is interpreted as a ``pinned" mass (constrained from motion), while the ``ON" state represents an unpinned mass (free to move).~Following UWCA rules, a monatomic square lattice with all masses at the ``OFF" state (i.e., pinned) is the starting point, and the mass at the lattice's center is then activated (i.e., unpinned pertaining to an ``ON" state).~Moving forward with an increasing number of UWCA generations, interesting patterns grow, adding more degrees of freedom after each generation due to the increasing number of unpinned masses.~Schematics of the first seven generations of the UWCA lattice are depicted in Figure~\ref{fig:Schematic}.

%\subsection{Mathematical model

\begin{figure}[H]
     \centering
\includegraphics[]{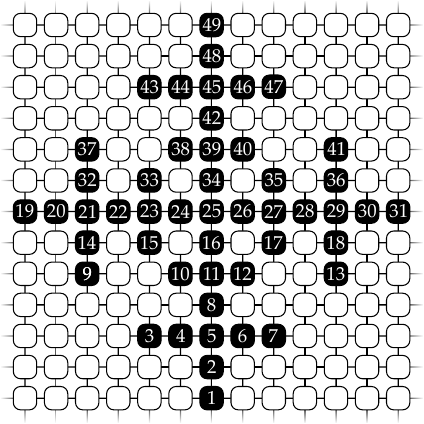}
\caption{\HBA{\textbf{Degrees of freedom numbering of the seventh generation UWCA lattice:} This schematic illustrates the degrees of freedom order for the seventh generation of the UWCA lattice, which comprises a total of forty-nine degrees of freedom (i.e., $n = 49$).}}
\label{fig:Numbering}
\end{figure}

To understand the UWCA lattice dynamics, the natural modes are computed for different generations.~To do so, I start by assembling the stiffness and mass matrices of a UWCA lattice at a given generation.~Denoting $n$ as the lattice's degrees of freedom, the equations of motion can be written in the matrix form as:
\begin{equation}
    \mathbf{M} \ddot{\mathbf{u}}+ \mathbf{K} {\mathbf{u}} = \mathbf{f}
    \label{eq:EOM}
\end{equation}
where $\mathbf{M}$ and $\mathbf{K}$ are the mass and stiffness matrices, respectively, while
\begin{subequations}
  \begin{equation}
    \mathbf{u} = 
    \begin{Bmatrix}
        u_1 & u_2 & \dots & u_n
    \end{Bmatrix}^{\text{T}}
\end{equation}
\begin{equation}
    \mathbf{f} = 
    \begin{Bmatrix}
        f_1 & f_2 & \dots & f_n
    \end{Bmatrix}^{\text{T}}
\end{equation}  
\end{subequations}
are the displacement and the forcing vectors, respectively. \HBA{The numbering of degrees of freedom commences from the very bottom row, and then from the first mass positioned on the far left of each row. This numbering convention persists throughout subsequent rows until the final top row.~A schematic illustrating the degrees of freedom numbering for the seventh generation of the UWCA lattice is provided in Figure~\ref{fig:Numbering}.} 

Given the discrete nature of the lattices, and uniform values of masses, the mass matrix is a constant diagonal matrix $\mathbf{M} = m\mathbf{I}$, where $\mathbf{I}$ is an identity matrix of size $n \times n$.~The stiffness matrix $\mathbf{K}$ is assembled according to the connectivity of the masses, which depends on the generation of UWCA. A common characteristic of $\mathbf{K}$ in all generations of UWCA lattice is the constant diagonal elements of $4k$, which will facilitate the discussion on its eigenvalues shortly.~Introducing $\mathbf{M}^{-1} \mathbf{K} = \omega_0^2 \mathbf{D}$, where $\omega_0 = \sqrt{k/m}$, and applying the harmonic solution in the absence of external forces, Equation~(\ref{eq:EOM}) reduces to:
\begin{equation}
\mathbf{D} \mathbf{u} = \lambda \mathbf{u}
\label{eq:EVP}
\end{equation}
which is a standard eigenvalue problem with eigenvalues $\lambda = \omega^2/\omega_0^2$ and $\omega$ being the frequency.~Examples of the matrix $\mathbf{D}$ corresponding to the seventh, eleventh, fifteenth, and nineteenth generations are provided in the supplementary materials.

%\subsection{Eigenfrequencies}

\begin{figure*}[]
     \centering
\includegraphics[]{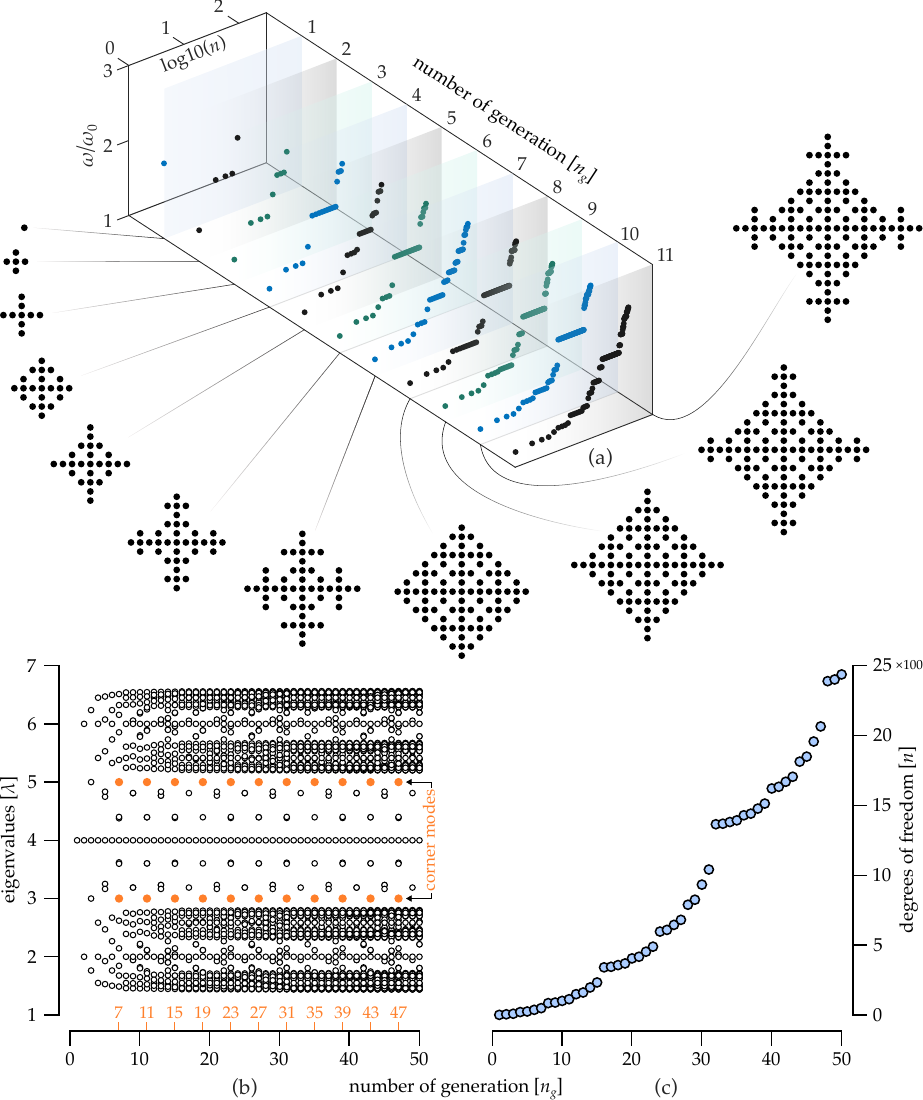}
\caption{\textbf{Eigenvalue analysis and growth of UWCA Lattices:} (a) The natural frequency spectra of elastic lattices constructed based on the first eleven generations of UWCA. A schematic of the free (or ``ON") masses for each generation is depicted for reference.~(b) The eigenvalue spectra of the UWCA lattice for an increasing number of generations, exhibiting symmetry about the eigenvalue $\lambda = 4$.~Among those eigenvalues, there are corner modes (colored in orange), which occur at odd numbers of generations, starting at seven and increasing in increments of four. (c) The degrees of freedom (denoted $n$) versus the number of generation (denoted $n_g$) is presented, demonstrating a considerable increase in the degrees of freedom as UWCA pattern grows with higher generations.}
\label{fig:Spectra}
\end{figure*}

Solving the eigenvalue problem in Equation~(\ref{eq:EVP}) using MATLAB, the eigenvalues (and thus the natural frequencies) are calculated for different generations $n_g$, as shown in Figure~\ref{fig:Spectra}(a) for the first eleven generations. As seen in the figure, the results are depicted as a three-dimensional plot, where the axes denote the natural frequency (normalized in units of $\omega_0$), the number of degrees of freedom ($n$) in a logarithmic scale of base 10, and the number of generation ($n_g$).~It is observed that the natural frequencies are bounded between $\omega/\omega_0 \approx 1.2$ and $\omega/\omega_0 \approx 2.6$, owing to the discrete nature of the lattice.~While an upper cutoff frequency is a known characteristic of discrete lattices, the lower cutoff frequency in the UWCA lattice is a byproduct of pinning, thus resembling a lattice with an elastic foundation~\cite{AlBabaa2017PoleDynamics,al2021enabling,nandi2020uncertainty}. Peculiarly, a signature of UWCA lattice dynamics is the repeated natural frequencies (the flat points in the figure), emerging at various frequencies (particularly at $\omega/\omega_0 = 2$), and their multiplicity depends on the overall shape of the structure at a given generation. 

If the eigenvalues $\lambda$ are plotted versus the number of generation $n_g$, interesting patterns are noted, as in Figure~\ref{fig:Spectra}(b).~Particularly, all eigenvalues are distributed symmetrically about $\lambda = 4$.~This symmetty becomes rather clear when the eigenvalues are shifted down by 4, given that matrix $\mathbf{D}$ has diagonal elements of identical magnitude, stemming from the constant diagonal elements of $\mathbf{K}$.~This process results in a spectrum where every eigenvalue $\lambda$ has a negative counterpart $-\lambda$, and being symmetric about zero instead of 4.~Such symmetry in eigenvalues is akin to particle-hole symmetry in condensed matter systems~\cite{Prodan2017DynamicalSystems}.~Finally, the degrees of freedom $n$ versus the number of generation $n_g$ is depicted in Figure~\ref{fig:Spectra}(c), showing a substantial increase in the degrees of freedom as UWCA pattern grows with higher generations. 

%\subsection{Corner Modes}
An interesting characteristic of a specific group of the repeated natural frequencies observed in Figures~\ref{fig:Spectra}(a,b) is the strong localization of deformation at the lattice's corners, known as corner modes, which have witnessed a spurt in research interest recently~\cite{serra2018observation,xue2019acoustic,Fan2019ElasticStates,danawe2021existence,al2024zero}.~These strongly-localized corner modes in UWCA lattices emerge at $\omega/\omega_0 = \sqrt{3}$ and $\omega/\omega_0 = \sqrt{5}$ (or $\lambda = 3$ and $\lambda = 5$) and only materialize when the generation number is according to the following formula: 
\begin{equation}
 n_g = 4r-1   
 \label{eq:ng}
\end{equation}
where $r$ is an integer equal or larger than 2.~As such, the number of generations $n_g$ having these special corner modes must be odd integers that start with $n_g = 7$ at $r=2$ and increase by an increment of four hereafter. At such generations, the UWCA lattice has corners with a perpendicular-sign ($\perp$) shape, such that six masses stem equally in three directions.~Each physical corner of the described shape corresponds to two repeated natural modes per frequency (i.e., two modes for each $\omega/\omega_0 = \sqrt{3}$ and~$\sqrt{5}$). \HBA{Finally, it is important to note that the third generation (i.e., $r=1$) exhibits four merged $\perp$ shapes, forming a lattice with an overall configuration resembling a plus sign ($+$), as shown in Figure~\ref{fig:Schematic}.~Consequently, the resulting structure have mode shapes with non-localized deformation at the corners, thus cannot be classified as corner modes.}
\begin{figure*}[!ht]
     \centering
\includegraphics[]{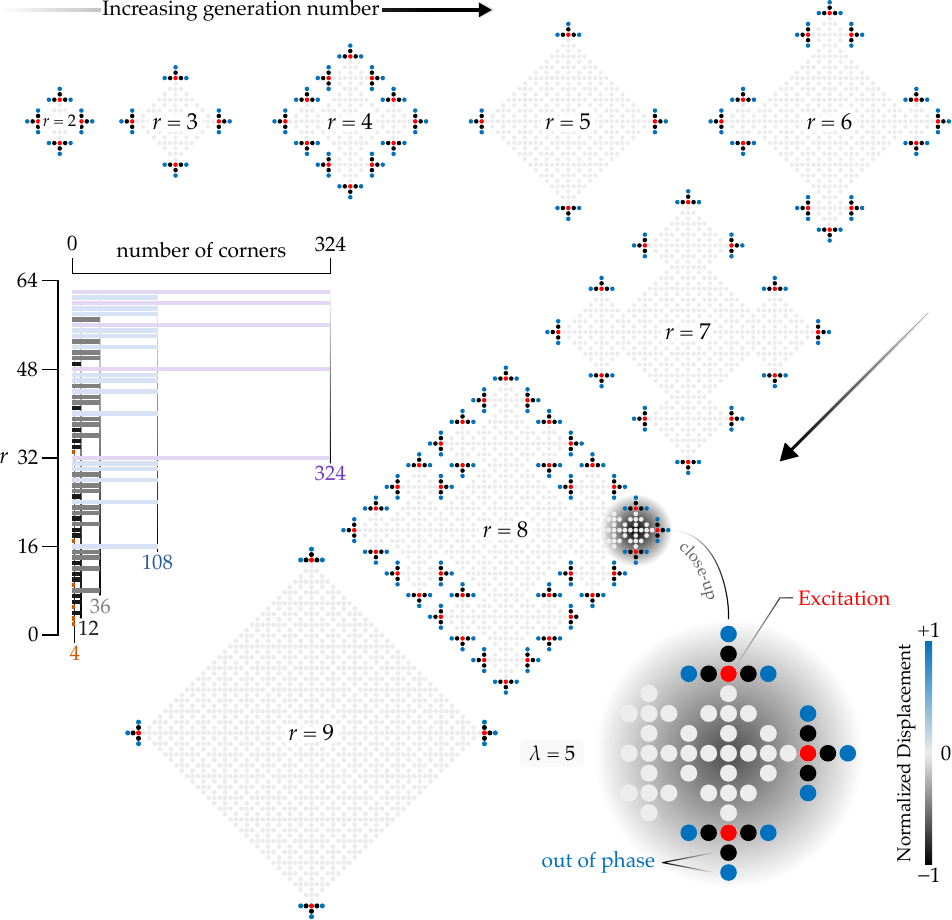}
\caption{\textbf{Corner modes in UWCA lattices:}~The response of the first eight generations of UWCA lattices with corners of $\perp$ shape, thus exhibiting strongly-localized corner modes, are shown for a frequency near $\omega = \sqrt{5} \omega_0$ (or $\lambda = 5$).~Corner modes are excited by applying a force at the junction connecting the $\perp$ shaped corner to the bulk of the lattice. These junctions are colored as red for ease of reference, and the normalized displacement response is indicated by the color bar.~The accompanying bar chart in the middle of the figure shows the number of corners of $\perp$ shape versus the integer~$r$. The number of corners differs with various values of $r$ (and generation number $n_g = 4r-1$), but they reset to four whenever $r = 2^z+1$ following a local maximum at $r = 2^z$, where $z$ is an integer larger than 1.}
\label{fig:Corner_modes}
\end{figure*}

In Figure~\ref{fig:Corner_modes}, the displacement response of UWCA lattice near $\omega/\omega_0 = \sqrt{5}$ (to avoid numerical singularity at $\sqrt{5}$) is calculated for eight successive generations having corner modes starting from $r = 2$.~The normalized magnitude of the displacement is reflected by the color legend.~To excite the corner modes, a harmonic force is applied at the conjunction of the $\perp$ shapes with the lattice's bulk, and are marked as red circles for ease of reference.~As can be seen in the figure, the bulk is nearly motionless, while the corner masses respond to the harmonic excitation.~Regardless of the value of integer $r$, the corner masses oscillate out of phase, as illustrated in the close-up of one of the corners in $r=8$ case.~If the excitation is near the other natural frequency that exhibits strongly-localized corner modes instead, i.e., $\omega/\omega_0 = \sqrt{3}$, then the corner masses shall move in phase.

It is crucial to shed light on the number of corners of $\perp$ shape with different values of $r$. In the middle of Figure~\ref{fig:Corner_modes}, the number of corners for a swept range of $r$ is depicted, for which the following observations are drawn:

\begin{enumerate}

\item The number of corners does not necessarily increase with increasing $r$. 

\item Whenever $r = 2^z + 1$, with $z$ being an integer larger than 1, the number of corners resets to four.~Observe, for instance, $r = 5$ and $r = 9$ cases (corresponding to $z = 2$ and $z = 3$, respectively), which are the successive integers of $r = 2^2 = 4$ and $r = 2^3 = 8$, respectively. 

\item Following the previous point, the number of corners locally maximizes at $r = 2^z$, immediately before it abruptly drops to four at the subsequent integer, i.e., at $r = 2^z + 1$.~Note that the corresponding generation numbers can be calculated via Equation~(\ref{eq:ng}).

    \item It is worth mentioning that the local maximum number of corners seems to grow with increasing~$z$. Each new maximum appears to triple the previous one as deduced from the sequence 4, 12, 36, 108, and 324 in the figure (i.e.,~$r =2^1, 2^2, 2^3, 2^4, \text{and}, 2^5$, respectively). 
            
\end{enumerate}

%\section{Concluding Remarks}
In summary, this paper introduced a new class of aperiodic lattices inspired by Ulam-Warburton Cellular Automaton (UWCA). The algorithm of UWCA starts with infinite packed squares that are initially at an ``OFF" state, and the squares are then activated to ``ON" following specific rules.~Such an algorithm results in aperiodic structures of an increasing number of ``ON" squares that grow larger with higher numbers of generations. Identical procedure is applied here to an infinite elastic monatomic lattice by interpreting the ``OFF" (``ON") states as pinned (unpinned) masses.~The eigenfrequency spectra of lattices constructed using the UWCA algorithm reveal unique dynamical properties, specifically the existence of strongly-localized corner modes.

Future directions of this research can include (1) exploring various types of CA algorithms for unique lattice designs and characteristics, (2) the study of damping characteristics to observe whether these lattice structures give rise to amplified damping (e.g., metadamping \cite{bacquet2018metadamping}), (3) investigating the effective elastic and inertial properties of these novel lattices, since lattices are generally known to be unique in that regard \cite{gonella2020symmetry}, (4) incorporating materials with tunable mechanical properties in their design (e.g., magnetoactive materials \cite{yu2018magnetoactive}), and (5) extending the same concept to acoustic lattices or LC-circuit networks.

\footnotesize
% \printbibliography[heading=none]
\bibliographystyle{IEEEtran}
\bibliography{references}
\end{multicols}
\end{document}